\documentclass[a4paper,12pt]{article}
\usepackage{amsmath}
\usepackage[round]{natbib}
\usepackage{array}
\usepackage{multirow}
\usepackage{booktabs}
\usepackage{graphicx}

\title{\textsc{Bayesian Estimation of Inverse\\ Gaussian Distribution}\thanks{This is an Author's Original Manuscript of an article submitted for consideration in the Journal of Statistical Computation \& Simulation \textcopyright Taylor \& Francis; Journal of Statistical Computation \& Simulation is available online at http://www.tandfonline.com/loi/gscs20}}
\author{\textsc{B. N. Pandey\thanks{Email: profbnpandey@yahoo.co.in}\, \& Pulastya Bandyopadhyay\thanks{Email: pulastya@gmail.com}}\\\textit{Department of Statistics, Banaras Hindu University,}\\\textit{Varanasi, India}}
\date{}

\begin{document}

\maketitle

\begin{abstract}
\noindent In this paper we consider Bayesian estimation for the parameters of inverse Gaussian distribution. Our emphasis is on Markov Chain Monte Carlo methods. We provide complete implementation of the Gibbs sampler algorithm. Assuming an informative prior, Bayes estimates are computed using the output of the Gibbs sampler and also from Lindley's approximation method. Maximum Likelihood and Uniformly Minimum Variance Unbiased estimates are obtained as well. We also compute Highest Posterior Density credible intervals, exact confidence intervals as well as \lq\lq percentile\rq\rq and \lq\lq percentile-t\rq\rq bootstrap approximations to the exact intervals. A simulation study was conducted to compare the long-run performance of the various point and interval estimation methods considered. One real data illustration has been provided which brings out some salient features of sampling-based approach to inference.
\end{abstract}



\section{Introduction}
\label{sec:1}
The inverse Gaussian (IG) distribution arises as the first passage time distribution in a Brownian motion process with positive drift. \citet{Tweedie57} first studied its basic characteristics and important statistical properties and showed certain analogies between its statistical analysis and that of the normal distribution. Chhikara and Folks explored the sampling theory inference for the distribution in great detail over several publications culminating in a review paper \citep{Folks78}. They also proposed it as a lifetime model and suggested its application in situations where the initial failure rate is high \citep{Chhikara77}. \citet{Johnson94} lists many useful applications of IG distribution in diverse fields ranging from theoretical physics to meteorology, reliability and lifetime data analysis, sequential analysis and industrial quality control, business applications and so on. The probability density function of the IG distribution is given by:
\begin{equation}
f(x|\,\mu,\lambda)=\displaystyle{\bigg(\frac{\lambda}{2\pi x^3}\bigg)}^{1/2}\exp\bigg[-\frac{\lambda (x-\mu)^2}{2\mu^2x}\bigg];\quad x>0,\,\mu,\lambda>0,
\label{eq:1}
\end{equation}
which we will denote by IG$(\mu,\lambda)$. The mean and the variance of the distribution is given by $E(X)=\mu$ and $Var(X)=\mu^3/\lambda$ respectively. Hence, neither $\mu$ can be interpreted as location nor $\lambda$ as scale parameter in the usual sense.

For a random sample $X_1,X_2,\dots,X_n$ from IG$(\mu,\lambda)$, the maximum likelihood estimators (MLE) of $\mu$ and $\lambda$ are given by:
\begin{equation}
\hat{\mu}=\bar{X} \quad\textrm{and}\quad \hat{\lambda}=n\bigg/\sum\limits_{i=1}^{n}(1/X_i-1/\bar{X}),
\label{eq:2}
\end{equation}
where $\bar{X}=\sum\limits_{i=1}^{n}X_i/n$, and the uniformly minimum variance unbiased estimators (UMVUE) are given by:
\begin{equation}
\tilde{\mu}=\hat{\mu}=\bar{X} \quad\textrm{and}\quad \tilde{\lambda}=(n-3)\bigg/\sum\limits_{i=1}^{n}(1/X_i-1/\bar{X}).
\label{eq:3}
\end{equation}

Bayesian analysis of the IG distribution has received considerable attention in the literature. \citet{Palmer73} first considered Bayesian estimation for the IG$(\mu,\lambda)$ model. He showed that when both parameters are treated as unknown, no natural conjugate prior exists and results become very difficult to obtain. This led to the use of other parametrizations of IG distribution in Bayesian studies. \citet{Banerjee79} worked with a parametrization involving $\psi=1/\mu$ and $\lambda$ and derived some Bayesian results with non-informative reference prior as well as the natural conjugate prior. Later works on the $(\psi,\lambda)$ parametrization include analysis based on Gibbs sampler and a Sampling Importance Resampling (SIR) scheme, assuming non-informative prior \citep{Shastri94} and approximate Bayesian estimation using natural conjugate prior \citep{Ahmad95}. \citet{Betro91} pointed out that in Banerjee and Bhattacharya's approach the posterior mean of $1/\psi$ does not exist, that is, the Bayes estimate of the mean of the distribution is not available. They chose another parametrization involving $\mu$ and $\phi$, where $\phi^{-1/2}$ is the coefficient of variation, arguing that it renders assessing the priors more natural because of the physical meaning attached to the parameters. They considered the marginal prior of $\phi$ to be a gamma density and the conditional prior of $\mu$ given $\phi$ to be an IG density and obtained Bayes estimates of the parameters as well as of their inverses.

Among the works on IG$(\mu,\lambda)$ parametrization, \citet{Padgett81} considered the estimation of reliability function. He noted that the Jeffreys' prior $\pi(\mu,\lambda)\propto(\mu^3\lambda)^{-1/2}$ does not lead to a proper posterior distribution and using the locally uniform non-informative reference prior $\pi(\mu|\,\lambda)\propto$ constant and $\pi(\lambda)\propto\lambda^{-1}$ found it extremely difficult to compute the estimate of reliability. As a workaround, he suggested a modified estimator. \citet{Sinha86} solved this problem by using Lindley's approximation technique. \citet{Ismail06} obtained estimates for $\lambda$ and its posterior density when $\mu$ is known, via Gibbs sampling, assuming Jeffreys' prior for $\lambda$.

In this paper, we focus on the IG$(\mu,\lambda)$ parametrization and provide full Bayesian analysis for the unrestricted model. As a substantial body of work is available for the $(\mu,\lambda)$ parametrization, especially in the classical framework, we believe it will be beneficial to stick to this representation as much as possible --- a belief also expressed in \citet{Palmer73,Shastri94}. We compute the Bayes estimates of the parameters under the assumption of informative priors. If quality prior information is available, then it is preferable to use informative priors rather than non-informative ones \citep[see][]{Berger85}.

Our emphasis here is on implementation of Markov Chain Monte Carlo (MCMC)-based computational methods. We provide complete posterior analysis of IG$(\mu,\lambda)$ distribution via Gibbs sampler. To put things in perspective, we employ different methods of estimation and see how they perform compared to each other. Besides computing the Bayes estimates of the parameters using the output of a Gibbs sampler, we also compute approximate Bayes estimates using Lindley's method. We compare the performance of the Bayes estimators with their classical counterparts, namely the MLEs and the UMVUEs given by Equation~(\ref{eq:2}) and Equation~(\ref{eq:3}), in terms of their mean squared error (MSE) by extensive simulation. Although, the use of frequentist risk analysis could be suspect from a purely Bayesian point of view, there are convincing arguments in favour of evaluating long run performance of a Bayesian procedure \citep[see][Section~4.8]{Berger85}. It is observed that the proposed Bayes estimators outperform the classical estimators.

We also consider different methods of interval estimation and compare the performance of the 95\% HPD credible intervals --- computed from the output of the Gibbs sampler --- with the 95\% exact confidence intervals and two different bootstrap approximations to the exact intervals -- the \lq\lq percentile\rq\rq bootstrap and \lq\lq percentile-t\rq\rq bootstrap intervals. Here also, the Bayesian credible intervals emerge as better.

The plan of the rest of the paper is as follows. In Section~\ref{sec:2}, we give the prior and the posterior distributions. Bayes estimation using Lindley's method and Gibbs sampler is outlined in Section~\ref{sec:3}. Setion~\ref{sec:4} contains different interval estimation methods discussed in detail. The simulation study is described and results are presented in Section~\ref{sec:5}. In Section~\ref{sec:6}, one real life dataset is analysed to illustrate our methodology. We summarize our findings and conclude the paper in Section~\ref{sec:7}.

\section{Prior and Posterior Distribution}
\label{sec:2}
Let ${\bf x}=(x_1,x_2,\dots,x_n)$ be a random sample of size $n$ from IG$(\mu,\lambda)$. The likelihood function of the observed sample can be written as:
\begin{equation}
L({\bf x};\,\mu,\lambda)=\displaystyle{\left(\frac{\lambda}{2\pi}\right)^{n/2}}\prod\limits_{i=1}^{n}x_i^{3/2}\exp\left[-\lambda(\alpha\mu^{-2}-n\mu^{-1}+\beta)\right],
\label{eq:4}
\end{equation}
where
\begin{equation}
\alpha=\sum\limits_{i=1}^{n}x_i/2\quad{\rm and}\quad\beta=\sum\limits_{i=1}^{n}x_i^{-1}/2
\label{eq:5}
\end{equation}
and $(\alpha,\beta)$ are jointly sufficient for $(\mu,\lambda)$.

Under the assumption of independence of $\mu$ and $\lambda$, the joint prior of $(\mu,\lambda)$ is $\pi(\mu,\lambda)=\pi_1(\mu)\,\pi_2(\lambda)$, where the prior for $\mu$ is a gamma density given by
\begin{equation}
\pi_1(\mu)=\displaystyle{\frac{b^a}{\Gamma(a)}}\mu^{a-1}\exp[-b\mu];\quad a,b>0,
\label{eq:6}
\end{equation}
and an independent prior for $\lambda$ is given by another gamma density
\begin{equation}
\pi_2(\lambda)=\displaystyle{\frac{d^c}{\Gamma(c)}}\lambda^{c-1}\exp[-d\lambda];\quad c,d>0.
\label{eq:7}
\end{equation}
Then the joint posterior of $(\mu,\lambda)$ can be written as
\begin{equation}
\rho(\mu,\lambda|\,{\bf x})=K\mu^{a-1}\exp[-b\mu]\,\lambda^{c+\frac{n}{2}-1}\exp\left[-\lambda(\alpha\mu^{-2}-n\mu^{-1}+\beta+d)\right],
\label{eq:8}
\end{equation}
where $K$ is the normalizing constant. Therefore, the Bayes estimate of $h(\mu,\lambda)$, any function of $\mu$ and $\lambda$, under squared error loss is the posterior expectation $\hat{h}_B(\mu,\lambda)=\int_{0}^{\infty}\int_{0}^{\infty}h(\mu,\lambda)\,\rho(\mu,\lambda|\,{\bf x})\,{\rm d}\mu \,{\rm d}\lambda$. This integral can not be computed in an explicit form. Hence, we discuss two different methods to compute the Bayes estimate in the next section.

\section{Bayesian Point Estimation}
\label{sec:3}
In Bayesian Statistics the posterior expectations frequently involve integrals which can not be computed explicitly. Various strategies to tackle this problem are available, e.g. adaptive quadrature based on numerical analysis, analytical approximation methods and sampling-based methods like Monte Carlo importance sampling and MCMC methods. Here, we  employ one approximate Bayes estimation method and one particular MCMC technique to compute the Bayes estimates of the parameters.

\subsection{Lindley's Approximation}
\label{sec:3.1}
\citet{Lindley80} suggested an asymptotic approximation to compute the ratio of two integrals. Using the prior mentioned in Section~\ref{sec:2}, the approximate Bayes estimates of $\mu$ and $\lambda$ under the squared error loss function, given by Lindley's method, turn out to be
\begin{equation}
\hat{\mu}_L=\hat{\mu}+(a+2)\frac{\hat{\mu}^2}{n\hat{\lambda}}-\frac{b\hat{\mu}^3}{n\hat{\lambda}}
\label{eq:9}
\end{equation}
and
\begin{equation}
\hat{\lambda}_L=\hat{\lambda}+(2c-1)\frac{\hat{\lambda}}{n}-\frac{2d\hat{\lambda}^2}{n},
\label{eq:10}
\end{equation}
where $\hat{\mu}$ and $\hat{\lambda}$ are the MLEs of $\mu$ and $\lambda$ as given in Equation~(\ref{eq:2}). We note that as $n\rightarrow\infty $, $\hat{\mu}_L\rightarrow\hat{\mu}$ and $\hat{\lambda}_L\rightarrow\hat{\lambda}$. The derivations of Equation~(\ref{eq:9}) and Equation~(\ref{eq:10}) are given in Appendix A.

Lindley's method is designed to compute the approximate Bayes estimates. MCMC methods, e.g. Gibbs sampling, which is discussed next, allows us to compute the Bayes estimates as well as to construct HPD credible intervals and summarize the marginal posterior distributions effectively using samples generated from the posterior.

\subsection{Gibbs Sampling}
\label{sec:3.2}
Gibbs sampling algorithm was first introduced by \citet{Geman84}, who applied the sampler on a Gibbs random field. A landmark paper by \citet{Gelfand90} first brought to attention its importance in general Bayesian computation. Today Gibbs sampler is the most widely used MCMC method and has found application in problems from diverse fields \citep[e.g.~for application in life testing and reliability as well as details of the algorithm see][]{Upadhyay01}.

From Equation~(\ref{eq:8}), the full conditionals for $\lambda$ and $\mu$ can be written as
\begin{align}
\label{eq:11}
\pi(\lambda|\,\mu,{\bf x})&\propto\lambda^{c+\frac{n}{2}-1}\exp\left[-\lambda(\alpha\mu^{-2}-n\mu^{-1}+\beta+d)\right]\\
\pi(\mu|\,\lambda,{\bf x})&\propto\mu^{a-1}\exp\left[-\alpha\lambda\mu^{-2}+n\lambda\mu^{-1}-b\mu\right].
\label{eq:12}
\end{align}
We see Equation~(\ref{eq:11}) is a gamma distribution and hence $\lambda$ can be easily generated using standard available generators. Generating $\mu$ from Equation~(\ref{eq:12}) can be a little tricky as Equation~(\ref{eq:12}) is not log-concave in $\mu$. We provide a procedure in Appendix B, which can be seen as an extension of the inverse transform method, which we use to generate samples from Equation~(\ref{eq:12}). This method was tested for several values of the parameters and it was found to be quite efficient.

Now that we have a mechanism to simulate Equation~(\ref{eq:11}) and Equation~(\ref{eq:12}), the posterior in Equation~(\ref{eq:8}) can be easily simulated via a two-stage Gibbs sampler. We took single long run of the chain and used ergodic averages for convergence monitoring. After discarding the initial transient phase, we picked up equally spaced outcomes from the remaining realizations of $(\mu,\lambda)$ and regarded them as samples from the true posterior. The spacing was so chosen that it renders the serial correlation negligible.

Then the Bayes estimates of $\mu$ and $\lambda$ under squared error loss function are given by
\begin{equation}
\hat{\mu}_B=\hat{E}(\mu|\,{\bf x})=\frac{1}{N}\sum\limits_{j=1}^{N}\mu_j\quad\textrm{and}\quad\hat{\lambda}_B=\hat{E}(\lambda|\,{\bf x})=\frac{1}{N}\sum\limits_{j=1}^{N}\lambda_j,
\label{eq:13}
\end{equation}
where $(\mu_j,\lambda_j)$ is the $j$-th MCMC sample, $j=1,2,\dots,N$.

\section{Interval Estimation}
\label{sec:4}
An interval estimate is often more useful than simply a point estimate. Taken together, the point estimate and the interval estimate say what is our best guess for the parameter value and how far in error that guess might reasonably be. In this section we describe four interval estimation procedures for $(\mu,\lambda)$.

\subsection{HPD Intervals}
\label{sec:4.1}
To construct HPD intervals for $\mu$ and $\lambda$ we follow the Monte Carlo procedure proposed by \citet{Chen99}. Given an MCMC sample $(\mu_j,\lambda_j)$; $j=1,2,\dots,N$, we calculate the HPD interval for $\mu$ as follows:\\
\textbf{Step 1.} Sort $\{\mu_j,j=1,2,\dots,N\}$ to obtain the ordered values
\[\mu_{(1)}\leq\mu_{(2)}\leq\dots\leq\mu_{(N)}.\]
\textbf{Step 2.} Compute the $100(1-\alpha)\%$ credible intervals
\[R_i(N)=(\mu_{(i)},\,\mu_{(i+[(1-\alpha)N])}),\quad\textrm{for}\quad i=1,2,\dots,N-[(1-\alpha)N],\]
where $[(1-\alpha)N]$ is the integer part of $(1-\alpha)N$.\\
\textbf{Step 3.} $R_{i^*}(N)$, the $100(1-\alpha)\%$ HPD interval, is the one with the smallest interval width among all $R_i(N)$'s.\\
The same procedure can be applied to calculate the HPD interval for $\lambda$.

\subsection{Exact Confidence Intervals}
\label{sec:4.2}
Confidence intervals for $\lambda$ parallel those for the variance of the normal distribution. A $100(1-\alpha)\%$ two sided confidence interval for $\lambda$ follows from the fact that $n\lambda V\sim\chi^2_{(n-1)}$, where $V=\sum\limits_{i=1}^{n}(1/X_i-1/\bar{X})/n=1/\hat{\lambda}$ \citep{Tweedie57}, and is given by
\begin{equation}
\left(\frac{\chi^2_{n-1,\,\alpha/2}}{nV},\,\frac{\chi^2_{n-1,\,1-\alpha/2}}{nV}\right),
\label{eq:14}
\end{equation}
where $\chi^2_{n-1,\,\alpha}$ is the $(100\cdot\alpha)$th percentile of $\chi^2$ distribution with $(n-1)$ degrees of freedom.

For unknown $\lambda$, a $100(1-\alpha)\%$ two sided confidence interval for $\mu$ was derived by \citet{Chhikara76} and is given by
\begin{equation}
\left(\bar{X}[1+\sqrt{\bar{X}V/(n-1)}\,t_{1-\alpha/2}]^{-1},\,\bar{X}[1-\sqrt{\bar{X}V/(n-1)}\,t_{1-\alpha/2}]^{-1}\right),
\label{eq:15}
\end{equation}
if $\left(1-\sqrt{\bar{X}V/(n-1)}\,t_{1-\alpha/2}\right)>0$, and $\left(\bar{X}[1+\sqrt{\bar{X}V/(n-1)}\,t_{1-\alpha/2}]^{-1},\,\infty\right)$, otherwise, where $t_{1-\alpha/2}$ is the $100(1-\alpha/2)$th percentile of Student's $t$ distribution with $(n-1) $ degrees of freedom.

\subsection{Bootstrap Confidence Intervals}
\label{sec:4.3}
In this subsection we discuss two bootstrap methods for constructing confidence intervals. We use the \lq\lq percentile\rq\rq or bootstrap-p method, which is the most widely used one and the \lq\lq percentile-t\rq\rq or bootstrap-t method, which is the most efficient one theoretically \citep[see][]{Efron82,Hall88}. Our constructions are based on the parametric bootstrap. We describe the methodology for $\mu$. Intervals for $\lambda$ can be constructed in a similar way.

The bootstrap-p confidence interval for $\mu$ can be obtained as follows:\\
\textbf{Step 1.} Calculate $\hat{\mu}$ and $\hat{\lambda}$ from ${\bf x}$.\\
\textbf{Step 2.} Generate $B$ bootstrap samples ${\bf x}\,_i^*$ of size $n$ from $IG(\hat{\mu},\hat{\lambda})$ and calculate $\hat{\mu}_i^*$ and $\hat{\lambda}_i^*$ from ${\bf x}\,_i^*$, $i=1,2,\dots,B$.\\
\textbf{Step 3.} The $100(1-\alpha)\%$ bootstrap-p confidence interval for $\mu$ is given by
\[\left(\hat{\mu}_B^{*\,(\alpha/2)},\,\hat{\mu}_B^{*\,(1-\alpha/2)}\right),\]
where $\hat{\mu}_B^{*(\alpha)}$ is the $(B\cdot\alpha)$th value in the ordered list of the $B$ replications of $\hat{\mu}^*$.

The bootstrap-t confidence interval for $\mu$ can be obtained as follows:\\
\textbf{Step 1.} Calculate $\hat{\mu}$ and $\hat{\lambda}$ from ${\bf x}$.\\
\textbf{Step 2.} Generate a bootstrap sample ${\bf x}^*$ of size $n$ from $IG(\hat{\mu},\hat{\lambda})$ and calculate $\hat{\mu}^*$ and $\hat{\lambda}^*$ from that.\\
\textbf{Step 3.} Generate a bootstrap sample ${\bf x}^{**}$ of size $n$ from $IG(\hat{\mu}^*,\hat{\lambda}^*)$ and calculate $\hat{\mu}^{**}$ and $\hat{\lambda}^{**}$ from that.\\
\textbf{Step 4.} Repeat Step 3 $B_2$ times.\\
\textbf{Step 5.} Calculate $se(\hat{\mu}^*)=\sqrt{\frac{1}{B_2-1}\sum\limits_{j=1}^{B_2}(\hat{\mu}_j^{**}-\bar{\mu}^{**})^ 2}$, where $\bar{\mu}^{**}=\sum\limits_{j=1}^{B_2}\hat{\mu}_j^{**}/B_2$.\\
\textbf{Step 6.} Calculate $T^*=\displaystyle{\frac{\hat{\mu}^*-\hat{\mu}}{se(\hat{\mu}^*)}}$.\\
\textbf{Step 7.} Repeat Steps 2-6 $B_1$ times.\\
\textbf{Step 8.} Calculate $se(\hat{\mu})=\sqrt{\frac{1}{B_2-1}\sum\limits_{j=1}^{B_2}(\hat{\mu}_j^*-\bar{\mu}^*)^ 2}$, where $\bar{\mu}^*=\sum\limits_{j=1}^{B_2}\hat{\mu}_j^*/B_2$.\\
\textbf{Step 9.} The $100(1-\alpha)\%$ bootstrap-t confidence interval for $\mu$ is given by
\[\left(\hat{\mu}-T_{B_1}^{*\,(1-\alpha/2)}\cdot se(\hat{\mu}),\,\hat{\mu}-T_{B_1}^{*\,(\alpha/2)}\cdot se(\hat{\mu})\right),\]
where $T_{B_1}^{*(\alpha)}$ is the $(B_1\cdot\alpha)$th value in the ordered list of the $B_1$ replications of $T^*$.

\section{Simulation Study}
\label{sec:5}
To compare the performance of the various estimation methods described above, a simulation study was performed. We took samples of size $n=15,20,30$ and $50$ from IG$(3,4)$. We assumed gamma$(6,2)$ as prior for $\mu$ and gamma$(5,1.25)$ as prior for $\lambda$, i.e. we chose $a=6,b=2,c=5$ and $d=1.25$. Under these priors, the Bayes estimates under squared error loss for $\mu$ and $\lambda$ were computed for each sample using both Lindley's method and the output of the Gibbs sampler. MCMC samples of size 1000 were taken for the computation. We also computed the MLEs and the UMVUEs. The average value of the estimates and their MSE, based on 1000 replications, are reported in Table~\ref{tab:1}.

The 95\% exact, bootstrap-p and bootstrap-t confidence intervals and HPD intervals for the parameters were also computed for each sample. For each interval given by $(l,u)$, we evaluated a shape factor measuring the skewness of the interval, which is $(u-\textrm{MLE})/(\textrm{MLE}-l)$ for the confidence intervals and $(u-\textrm{posterior mode})/(\textrm{posterior mode}-l)$ for the HPD intervals. We also calculated two factors as MissLeft and MissRight for each interval, where MissLeft implies that the true value of the parameter falls below $l$ and MissRight means that it lies above $u$. Table~\ref{tab:2} gives the average intervals along with their average shape and coverage, including the MissLeft and MissRight probabilities, based on 1000 replications.
\begin{table}[htbp]
\caption{Average and MSE (within parenthesis) of different estimators for different sample sizes. The first entry in each cell corresponds to $\mu$ and the second to $\lambda$.}
\scalebox{0.991}{\begin{tabular}{*{5}{c}}
\noalign{\smallskip}\toprule
$n$ &  \textbf{MLE} &  \textbf{UMVUE} &  \textbf{Bayes(Lindley)} & \textbf{Bayes(Gibbs)} \\ 
\midrule
\multirow{2}{*}{\textbf{15}} & 3.0028(0.4396) & 3.0028(0.4396) & 3.2103(0.3444) & 3.1426(0.2764) \\
  & 4.9167(5.4989) & 3.9333(2.9860) & 3.0613(12.3134) & 4.1909(0.8051) \\ 
\cmidrule(lr){2-5}
\multirow{2}{*}{\textbf{20}} & 3.0087(0.3212) & 3.0087(0.3212) & 3.1831(0.2799) & 3.1357(0.2343) \\
  & 4.6511(3.2504) & 3.9535(2.0442) & 3.6867(0.7075) & 4.1771(0.7859) \\ 
\cmidrule(lr){2-5}
\multirow{2}{*}{\textbf{30}} & 3.0076(0.2220) & 3.0076(0.2220) & 3.1327(0.2072) & 3.1106(0.1868) \\
  & 4.4509(1.6967) & 4.0059(1.2096) & 4.0109(0.3317) & 4.1758(0.6725) \\ 
\cmidrule(lr){2-5}
\multirow{2}{*}{\textbf{50}} & 2.9924(0.1363) & 2.9924(0.1363) & 3.0734(0.1312) & 3.0663(0.1260) \\
  & 4.2756(0.8509) & 4.0191(0.6851) & 4.0924(0.4033) & 4.1363(0.4879) \\ 
\bottomrule
\end{tabular}}
\label{tab:1}
\end{table}

\begin{table}[htbp]
\caption{Average 95\% intervals and their shape and coverage probabilities for different sample sizes. The first entry in each cell corresponds to $\mu$ and the second to $\lambda$.}
\scalebox{0.965}{\begin{tabular}{c>{\bfseries}c*{5}{c}}
\noalign{\smallskip}\toprule
 $n$ &  & Interval & Shape & Coverage & MissLeft & MissRight \\ 
\midrule
 \multirow{8}{*}{\textbf{15}} & Exact & (2.0359, 6.0797) & 2.9480 & 0.956 & 0.018 & 0.026 \\ 
  &  & (1.8450, 8.5612) & 1.1865 & 0.960 & 0.021 & 0.019 \\ 
\cmidrule(lr){3-7}
 & Boot-p & (1.9349, 4.4723) & 1.3519 & 0.911 & 0.006 & 0.083 \\ 
 & & (2.8191, 13.1053) & 3.9045 & 0.862 & 0.138 & 0.000 \\ 
\cmidrule(lr){3-7}
 & Boot-t & (2.0587, 5.4791) & 2.5083 & 0.944 & 0.023 & 0.033 \\ 
 & & (1.4864, 8.8460) & 1.1465 & 0.952 & 0.025 & 0.023  \\ 
\cmidrule(lr){3-7}
 & HPD & (2.0346, 4.4732) & 1.9159 & 0.973 & 0.008 &  0.019 \\ 
 & & (2.0341, 6.5653) & 1.5038 & 0.989 & 0.000 & 0.011  \\ 
\midrule
 \multirow{8}{*}{\textbf{20}} & Exact & (2.1441, 5.1679) & 2.3929 & 0.956 & 0.020 & 0.024 \\ 
  &  & (2.0713, 7.6400) & 1.1585 & 0.955 & 0.021 & 0.024  \\ 
\cmidrule(lr){3-7}
 & Boot-p & (2.0507, 4.2688) & 1.3004 & 0.927 & 0.008 &  0.065 \\ 
 & & (2.8284, 10.4286) & 3.1769 & 0.896 & 0.104 & 0.000  \\ 
\cmidrule(lr){3-7}
 & Boot-t & (2.1707, 4.9529) & 2.2531 & 0.950 & 0.023 & 0.027  \\ 
 & & (1.8732, 7.7919) & 1.1303 & 0.953 & 0.026 & 0.021  \\ 
\cmidrule(lr){3-7}
 & HPD & (2.1334, 4.3196) & 1.8339 & 0.974 & 0.008 &  0.018 \\ 
 & & (2.1918, 6.3284) & 1.4441 & 0.981 & 0.002 & 0.017  \\  
\midrule
 \multirow{8}{*}{\textbf{30}} & Exact & (2.2764, 4.4662) & 1.9528 & 0.953 & 0.020 & 0.027 \\ 
  &  & (2.3808, 6.7836) & 1.1268 & 0.951 & 0.032 & 0.017  \\ 
\cmidrule(lr){3-7}
 & Boot-p & (2.1987, 4.0202) & 1.2437 & 0.930 & 0.010 &  0.060 \\ 
 & & (2.9200, 8.3060) & 2.5194 & 0.905 & 0.095 & 0.000  \\ 
\cmidrule(lr){3-7}
 & Boot-t & (2.2905, 4.4334) & 1.9542 & 0.943 & 0.025 & 0.032  \\ 
 & & (2.2752, 6.8739) & 1.1154 & 0.953 & 0.026 & 0.021  \\ 
\cmidrule(lr){3-7}
 & HPD & (2.2586, 4.0954) & 1.7282 & 0.961 & 0.013 &  0.026 \\ 
 & & (2.4484, 6.0391) & 1.3764 & 0.981 & 0.004 & 0.015  \\  
 \midrule
  \multirow{8}{*}{\textbf{50}} & Exact & (2.4047, 3.9693) & 1.6465 & 0.951 & 0.019 & 0.030 \\ 
   &  & (2.6983, 6.0049) & 1.0964 & 0.947 & 0.030 & 0.023  \\ 
\cmidrule(lr){3-7}
  & Boot-p & (2.3468, 3.7587) & 1.1836 & 0.932 & 0.009 &  0.059 \\ 
  & & (3.0425, 6.7622) & 2.0179 & 0.921 & 0.077 & 0.002  \\ 
 \cmidrule(lr){3-7}
  & Boot-t & (2.4133, 3.9833) & 1.6970 & 0.939 & 0.022 & 0.039  \\ 
  & & (2.6452, 6.0585) & 1.0947 & 0.954 & 0.027 & 0.019  \\ 
 \cmidrule(lr){3-7}
  & HPD & (2.3877, 3.8254) & 1.5774 & 0.956 & 0.012 &  0.032 \\ 
  & & (2.7222, 5.6353) & 1.2881 & 0.968 & 0.012 & 0.020  \\  
\bottomrule
\end{tabular}}
\label{tab:2}
\end{table}

Table~\ref{tab:1} brings certain points to our attention. First of all, it is observed that as expected, the MSE of each estimator decreases as the sample size increases. Bayes estimators obtained from both Lindley's method and Gibbs sampler are observed to perform much better than the classical estimators, but the discrepancy in their relative performance tends to get smaller and smaller with the increase in sample size. We observe that except for estimation of $\lambda$ in small samples, Bayes estimators obtained from Lindley's method perform quite well. This can be explained by the rather erratic behaviour of the MLE of $\lambda$ in small samples. As the performance of the MLE of $\lambda$ gets stabilised with larger sample size, Lindley's method quickly catches up and performs much better for moderate to large sized samples. Overall, the Bayes estimators computed from our proposed Gibbs sampler are found to be performing very well.

From Table~\ref{tab:2} we see that as par our expectation, length of all the intervals decrease with increase in sample size. We also observe that the coverage probability of each interval remains quite close to the nominal value of 95\%. As the exact confidence interval, and hence the bootstrap intervals, are equi-tailed by construction, the ideal values of MissLeft and MissRight probabilities are 0.025 each. The bootstrap-t approximates the shape of the exact intervals very well and as a result gets both the endpoints and balance at both ends right. The bootstrap-p gets the tail area probabilities wrong for both MLEs and hence, as approximation of exact intervals, it fares poorly in comparison to bootstrap-t. It is seen that for all the intervals including the HPD, shape $\rightarrow 1$  with increase in $n$, i.e. the intervals become more and more symmetric. This results from the asymptotic normality of the underlying distributions of the MLEs and the posteriors of $\mu$ and $\lambda$. For symmetrical posteriors, the HPD interval becomes equi-tailed. Indeed, we observe that as sample size increases, the MissLeft and MissRight probabilities for the HPD get stabilised toward 0.025. Overall, we see that the HPD intervals work very well in terms of both length and coverage.

\section{Numerical Illustration}
\label{sec:6}
In this section, we consider a real data set initially analysed by \citet{Chhikara77}. The maintenance data set given in Table~\ref{tab:3} represents active repair times (in hours) for an airborne communication transceiver. 
\begin{table}[htbp]
\caption{Repair times (in hours) of 46 transceivers}
\centering
\begin{tabular}{*{10}{r}}
\noalign{\smallskip}\toprule
0.2 & 0.3 & 0.5 & 0.5 & 0.5 & 0.5 & 0.6 & 0.6 & 0.7 & 0.7 \\ 
0.7 & 0.8 & 0.8 & 1.0 & 1.0 & 1.0 & 1.0 & 1.1 & 1.3 & 1.5 \\ 
1.5 & 1.5 & 1.5 & 2.0 & 2.0 & 2.2 & 2.5 & 2.7 & 3.0 & 3.0 \\ 
3.3 & 3.3 & 4.0 & 4.0 & 4.5 & 4.7 & 5.0 & 5.4 & 5.4 & 7.0 \\ 
7.5 & 8.8 & 9.0 & 10.3 & 22.0 & 24.5 &  &  &  &  \\ 
 \bottomrule
\end{tabular}
\label{tab:3} 
\end{table}
Among the previous works done on this data set, \citet{Chhikara77} obtained unbiased and maximum likelihood estimation of IG parameters and showed that the IG model is a good fit to the data using Kolmogorov--Smirnov test. \citet{Sinha86} computed the Bayes estimates using Lindley's method under a non-informative prior. \citet{Betro91} also analysed this data set and calculated the estimates for the parameters.

\begin{figure}[htbp]
\centering
\includegraphics[width=0.8\textwidth]{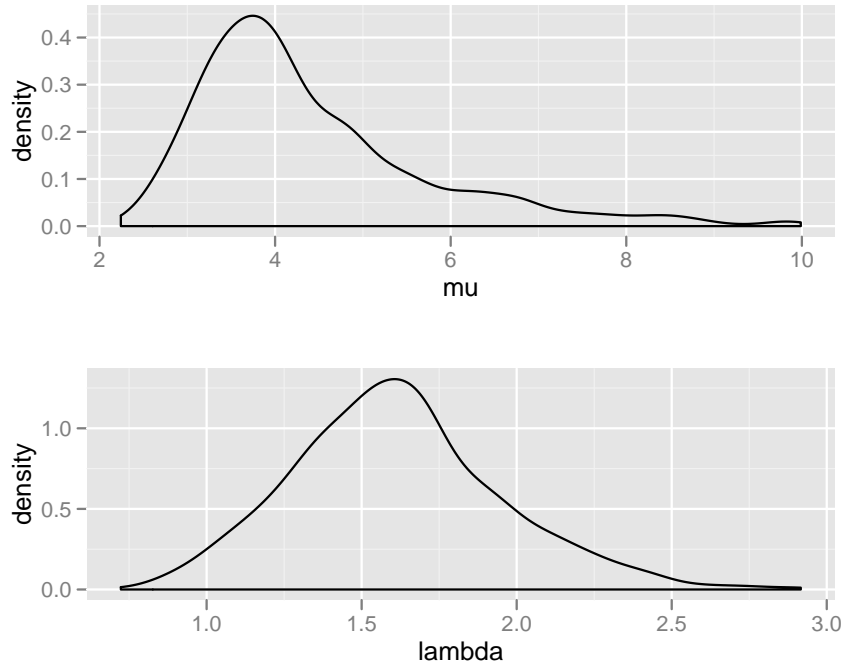}
\caption{Kernel density estimates for marginal posterior densities of $\mu$ and $\lambda$ for the repair times data}
\label{fig:1}
\end{figure}

For our analysis, in absence of prior information, we assumed the non-informative prior $\pi(\mu|\,\lambda)\propto$ constant and $\pi(\lambda)\propto\lambda^{-1}$, the same prior used by \citet{Padgett81} and \citet{Sinha86}. The joint posterior of $(\mu,\lambda)$ under this prior can be written as
\begin{equation}
\rho(\mu,\lambda|\,{\bf x})\propto\lambda^{\frac{n}{2}-1}\exp\left[-\lambda(\alpha\mu^{-2}-n\mu^{-1}+\beta)\right].
\label{eq:16}
\end{equation}
This posterior can be obtained by putting $a=1,b=0,c=0$ and $d=0$ in Equation~(\ref{eq:8}). So, for our analysis we could use the same Gibbs sampler as described in Section~\ref{sec:3.2}, without any further modification, for this particular choice of $a,b,c$ and $d$. We used MLE as the initial value for the chain and took MCMC samples of size 1000 to compute the Bayes estimates. Some sample based posterior characteristics are presented in Table~\ref{tab:4}. Figure~\ref{fig:1} depicts the marginal posterior density estimates of $\mu$ and $\lambda$ using the Gaussian kernel, based on the MCMC samples. The interval estimates obtained, are presented in Table~\ref{tab:5}.
\begin{table}[htbp]
\caption{Sample based posterior characteristics for the repair times data}
\scalebox{0.79}{\begin{tabular}{*{9}{c}}
\noalign{\smallskip}\toprule
 & Mean & Variance & Mode & 1st Quartile & Median & 3rd Quartile & Minimum & Maximum \\ 
 \midrule
$\mu$ & 4.3999 & 1.8245 & 3.7475 & 3.4750 & 4.0229 & 4.9173 & 2.2434 & 9.9877 \\ 
$\lambda$ & 1.6129 & 0.1164 & 1.6065 & 1.3791 & 1.5953 & 1.8092 & 0.7234 & 2.9163 \\
\bottomrule 
\end{tabular} }
\label{tab:4}
\end{table}

Using the vague prior mentioned above, \citet{Sinha86} obtained the expressions for the Bayes estimates of $\mu$ and $\lambda$, using Lindley's method, as:
\begin{equation}
\hat{\mu}_L=\hat{\mu}+\frac{3\hat{\mu}^2}{n\hat{\lambda}}\quad\textrm{and}\quad\hat{\lambda}_L=\left(\frac{n-1}{n}\right)\hat{\lambda},
\label{eq:17}
\end{equation}
which, again, can be derived from Equation~(\ref{eq:9}) and Equation~(\ref{eq:10}) for the same choice of $a,b,c$ and $d$ as above.

\begin{table}[tbp]
\caption{95\% intervals and their shape for the repair times data}
\scalebox{0.87}{\begin{tabular}{*{6}{c}}
\noalign{\smallskip}\toprule
 &  & Exact & Boot-p & Boot-t & HPD \\ 
\midrule
 \multirow{2}{*}{$\mu$} & Interval & (2.4998, 6.4715) & (2.2868, 5.4372) & (2.4251, 6.5312) & (2.4134, 7.2643) \\ 
  & Shape & 2.5888 & 1.3872 & 2.4756 & 2.6361 \\ 
\cmidrule(lr){3-6}
\multirow{2}{*}{$\lambda$}  & Interval & (1.0229, 2.3588) & (1.1950, 2.7510) & (0.8674, 2.4426) & (0.9268, 2.2686) \\ 
  & Shape & 1.1007 & 2.3547 & 0.9903 & 0.9742 \\ 
\bottomrule
\end{tabular}}
\label{tab:5}
\end{table}

The MLEs, UMVUEs and Bayes estimates from Lindley's method for the parameters, as obtained by \citet{Sinha86}, are $\hat{\mu}=3.6065$, $\hat{\lambda}=1.6589$; $\tilde{\mu}=3.6065$, $\tilde{\lambda}=1.5507$; $\hat{\mu}_L=4.1178$ and $\hat{\lambda}_L=1.6228$ respectively. The Bayes estimates obtained from our proposed Gibbs sampler turn out to be $\hat{\mu}_B=4.3999$ and $\hat{\lambda}_B=1.6129$. The Bayes estimate of $\mu$ provided by \citet{Betro91} is 3.653. So, the estimates obtained are in reasonable agreement with previously proposed estimates.

\begin{figure}[tbp]
\centering
\resizebox*{10cm}{!}{\includegraphics{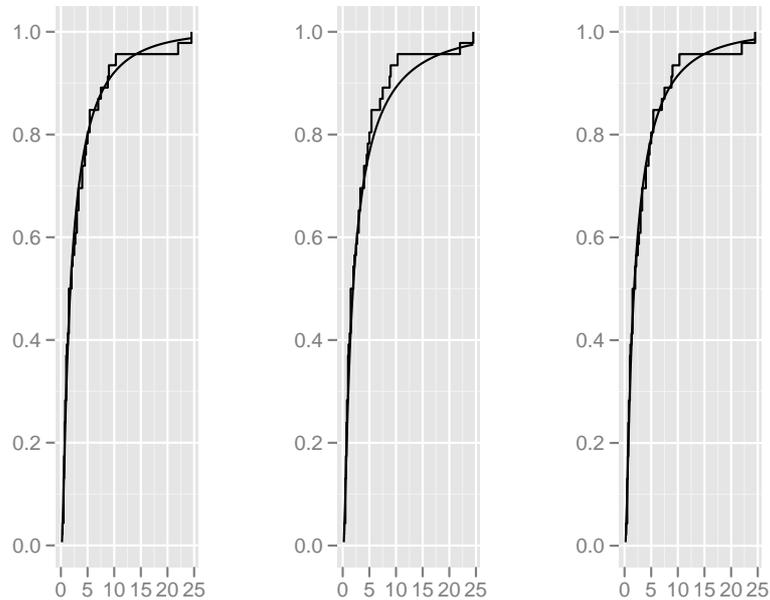}}
\caption{The empirical and fitted CDFs for the repair times data using different estimates: \emph{left}: MLE for $\mu$ and $\lambda$ \emph{centre}: posterior mean for $\mu$ and $\lambda$ \emph{right}: posterior mode for $\mu$ and posterior mean for $\lambda$}
\label{fig:2}
\end{figure}

Figure~\ref{fig:1} and Table~\ref{tab:4} reveals the advantage of the MCMC-based or sampling-based approach towards inference. Full posterior analysis based on MCMC samples, gives an overall view of the underlying posteriors and suggests the use of appropriate estimators for inference purpose. For example, for the repair times data, unlike in the case of $\lambda$, a little difference can be seen between possible estimates of $\mu$. Now, Figure~\ref{fig:1} shows that the marginal posterior of $\mu$ is much more skewed than that of $\lambda$. This suggests that for $\mu$, HPD intervals are more preferable than equi-tailed intervals and modal value might be a more reasonable point estimate. Indeed, from Figure~\ref{fig:2} we can see that using posterior mode instead of posterior mean for $\mu$, we can obtain a better fit to the data.

\section{Conclusion}
\label{sec:7}
In this paper, we have considered Bayesian estimation for the parameters of the IG$(\mu,\lambda)$ distribution. We assumed independent gamma priors for both the parameters in this study. Complete implementation of the Gibbs sampling algorithm has been provided. Simply by choosing suitable hyper-parameters, and without any further modification, the proposed Gibbs sampler is shown to work as smoothly for a particular non-informative prior, which has been used previously for studying the IG distribution. We have used squared error loss function in this paper, but the proposed methodology can be routinely applied for other asymmetric loss functions as well. We have shown how techniques based on MCMC can easily deal with the issue of awkward posterior, perform well compared to other available methods of estimation and on top of that, provide a conceptually simple, routinely implementable and unified framework for complete analysis of the IG distribution. But perhaps the greatest advantage of this sampling-based approach is the unique insight it offers us into the posteriors. This enables us to make better inference by helping us choose appropriate inference procedures, in particular when the posterior turns out to be non-normal and skewed.

\section*{Appendix A}
\label{app:a}
For a model involving two parameters $(\theta_1,\theta_2)$, Lindley's approximation technique \citep{Lindley80} gives an expression for the posterior expectation of $u$, a function of $(\theta_1,\theta_2)$, as:
\begin{align}
E(u|{\bf x})&=u+\frac{1}{2}\sum\limits_{i=1}^2\sum\limits_{j=1}^2u_{ij}\sigma_{ij}+\sum\limits_{j=1}^2U_j\rho_j+\frac{1}{2}L_{111}\sigma_{11}U_1\notag\\&+\frac{1}{2}L_{112}(2\sigma_{12}U_1+\sigma_{11}U_2)+\frac{1}{2}L_{122}(\sigma_{22}U_1+2\sigma_{12}U_2)\notag\\&+\frac{1}{2}L_{222}\sigma_{22}U_2,
\label{eq:18}
\end{align}
where all the terms are evaluated at the MLE $(\hat{\theta}_1,\hat{\theta}_2)$ and $U_k=\sum\limits_iu_i\sigma_{ki}$, the partial derivatives $\rho_j,u_j,u_{ij},L_{ijk}$ are defined as, for example, $L_{ijk}=\frac{\partial^3L}{\partial\theta_i\partial\theta_j\partial\theta_k}$, where each suffix denotes differentiation once with respect to the variable having that suffix, $\sigma_{ij}$ are the elements of the inverse of the matrix having elements $\{-L_{ij}\}$, $L$ is the log-likelihood and $\rho$ is the logarithm of the joint prior.

Now, considering $(\theta_1,\theta_2)=(\mu,\lambda)$, the definitions of likelihood function and priors given in Section~\ref{sec:2} lead to:
\begin{align*}
&L_{111}=\frac{6n\hat{\lambda}}{\hat{\mu}^4}, L_{112}=-\frac{n}{\hat{\mu}^3}, L_{122}=0, L_{222}=\frac{n}{\hat{\lambda}^3},\\&\sigma_{11}=\frac{\hat{\mu}^3}{n\hat{\lambda}}, \quad\sigma_{12}=\sigma_{21}=0,\quad\sigma_{22}=\frac{2\hat{\lambda}^2}{n},\\&\rho_1=\frac{a-1}{\hat{\mu}}-b\quad\textrm{and}\quad\rho_2=\frac{c-1}{\hat{\lambda}}-d.
\end{align*}
Considering  $u=\mu$ and $u=\lambda$ in turn and using the above values, from Equation~(\ref{eq:18}) we get the expressions of Bayes estimators given by Equation~(\ref{eq:9}) and Equation~(\ref{eq:10}) respectively.

\section*{Appendix B}
\label{app:b}
Let $X$ be a random variable having probability density function $f(\cdot)$ and cumulative distribution function $F(\cdot)$. Then $F(x)\sim U(0,1)$, and this result is the foundation of the inverse transform method of random variate generation. We generate a $u$ from $U(0,1)$, and find out $c$ such that $F(c)=u$. Then $c$ can be taken as a random sample from $f(\cdot)$.

Now, let $f(x)=kg(x)$, where $k$ is a constant, i.e. $g(\cdot)$ is the non-normalized density. Then, defining  $F^*(x)=\int_{-\infty}^{x}g(t)\textrm{d}t=F(x)/k$, we have $F^*(x)\sim U(0,1/k)$. So, generating $u$ from $U(0,1/k)$ and solving for $c$ in $F^*(c)=u$, also gives us a random sample $c$ from $f(\cdot)$.

We use this idea to generate $\mu$ from Equation~(\ref{eq:12}). First, we find a $c$ such that $\int_{0}^{c}\pi(\mu|\lambda,{\bf x})\textrm{d}\mu\approx\int_{0}^{\infty}\pi(\mu|\lambda,{\bf x})\textrm{d}\mu$, i.e. integrating $\pi(\mu|\lambda,{\bf x})$ upto $c$, yields almost --- for example, 99.9999\% of --- the whole area under the curve. Let $\int_{0}^{c}\pi(\mu|\lambda,{\bf x})\textrm{d}\mu=T$. We generate a $u$ from $U(0,T)$ and solve $\int_{0}^{\mu^*}\pi(\mu|\lambda,{\bf x})\textrm{d}\mu=u$ for $\mu^*$. Then $\mu^*$ is a random sample from $\pi(\mu|\lambda,{\bf x})$.

\bibliographystyle{newapa}
\bibliography{BayesianIG}

\end{document}